\documentclass[conference, 10pt]{IEEEtran}
\usepackage[dvips]{graphicx}
\usepackage{amssymb,amsmath}
\usepackage{algorithm2e}
\usepackage{subfig}
\IEEEoverridecommandlockouts
\newtheorem{theorem}{Theorem}[section]

\newtheorem{corollary}[theorem]{Corollary}

\newcommand{\ie}{\emph{i.e. }}

\begin{document}

\title{A Multi-hop Multi-source Algebraic Watchdog}
\author{
\authorblockN{MinJi Kim\authorrefmark{1}, Muriel M\'{e}dard\authorrefmark{1}, Jo\~{a}o Barros\authorrefmark{2}\vspace*{.2cm}}
\hspace*{2cm} \authorrefmark{1}{\normalsize Research Laboratory of Electronics} \hspace*{3.3cm} \authorrefmark{2}{\normalsize Instituto de Telecommunica\c{c}\~{o}es}\hfill\\
\hspace*{2cm} {\normalsize Massachusetts Institute of Technology} \hspace*{1cm} {\normalsize Departamento de Engenharia Electrot\'{e}cnica e de Computadores}\hfill\\
\hspace*{2.5cm}{\normalsize Cambridge, MA 02139, USA} \hspace*{2cm} {\normalsize Faculdade de Engenharia da Universidade do Porto, Portugal}\hfill\\
\hspace*{2.2cm}{\normalsize Email: \{minjikim, medard\}@mit.edu} \hspace*{4cm} {\normalsize Email: jbarros@fe.up.pt}\hfill\vspace*{-.5cm}
}

\maketitle

\begin{abstract}
In our previous work (`An Algebraic Watchdog for Wireless Network Coding'), we proposed a new scheme in which nodes can detect malicious behaviors probabilistically, police their downstream neighbors locally using overheard messages; thus, provide a secure global \emph{self-checking network}. As the first building block of such a system, we focused on a two-hop network, and presented a graphical model to understand the inference process by which nodes police their downstream neighbors and to compute the probabilities of misdetection and false detection.

In this paper, we extend the Algebraic Watchdog to a more general network setting, and propose a protocol in which we can establish \emph{trust} in coded systems in a distributed manner. We develop a graphical model to detect the presence of an adversarial node downstream within a general two-hop network. The structure of the graphical model (a trellis) lends itself to well-known algorithms, such as Viterbi algorithm, that can compute the probabilities of misdetection and false detection. Using this as a building block, we generalize our scheme to multi-hop networks. We show analytically that as long as the min-cut is not dominated by the Byzantine adversaries, upstream nodes can monitor downstream neighbors and allow reliable communication with certain probability. Finally, we present preliminary simulation results that support our analysis.
\end{abstract}
\IEEEpeerreviewmaketitle

\section{Introduction}

We consider the problem of Byzantine detection in a coded wireless network. Previous work on Byzantine detection focused on receiver-based protocols, in which the destination nodes of the corrupted data detects the presence of an adversary upstream. However, this detection may come too late as the adversary is partially successful in disrupting the network (even if it is detected). It has wasted network bandwidth, while the source is still unaware of the need for retransmission.

In our previous work \cite{algebraicwatchdog}, we proposed a new scheme called the \emph{algebraic watchdog}, in which nodes can detect malicious behaviors probabilistically by taking advantage of the broadcast nature of the wireless medium. The algebraic watchdog was inspired by the an analogous protocol for routing wireless networks, called the \emph{watchdog and pathrater} \cite{marti}. The key difference between the our previous work \cite{algebraicwatchdog} and that of \cite{marti} is that we allow network coding. Network coding \cite{ahlswede}\cite{algebraic} is advantageous as it increases throughput, is robust against failures/erasures, and is resilient in dynamic networks.

The key challenge in algebraic watchdog is that, by incorporating network coding, we can no longer recognize packets individually. In \cite{marti}, a node $v$ can monitor its downstream neighbor $v'$ by checking that the packet transmitted by $v'$ is a copy of what $v$ transmitted to $v'$. However, with network coding, this is no longer possible as transmitted packets are a function of the received packets. Furthermore, $v$ may not have full information regarding the packets received at $v'$; thus, node $v$ is faced with the challenge of inferring the packets received at $v'$ and ensuring that $v'$ is transmitting a valid function of the received packets. We note that \cite{whenmeets} combines source coding with watchdog; thus, \cite{whenmeets} does not face the same problem as that of algebraic watchdog.

\section{Problem Statement}\label{sec:problemstatement}
We use elements from a field, and their bit-representation. We use the same character in italic font (\ie $x$) for the field element, in bold font (\ie $\mathbf{x}$) for the bit-representation, and in underscore bold font (\ie $\mathbf{\underline{x}}$) for vectors.
For arithmetic operations in the field, we shall use the conventional notation (\ie $+, -, \cdot$).
For bit-wise addition, we use $\oplus$.

The problem statement for this paper is similar to that in our previous work \cite{algebraicwatchdog}. A wireless network is modeled using a directed graph $G = (V, E_1, E_2)$, where $V$ is the set of network nodes, $E_1$ the set of intended transmissions, and $E_2$ the set of interference channels. If $(v_i, v_j) \in E_1$ and $(v_i, v_k) \in E_2$ where $v_i, v_j, v_k \in V$, then there is an intended transmission from $v_i$ to $v_j$, and $v_k$ can overhear this transmission with noise (modeled using binary symmetric channel $BSC(p_{ik})$).
Node $v_i\in V$ transmits a coded packet $\mathbf{\underline{p_i}}$, where $\mathbf{\underline{p_i}}=[\mathbf{a_i}, \mathbf{h_{I_i}}, \mathbf{h_{x_i}}, \mathbf{x_i}]$ is a $\{0,1\}$-vector. A valid packet $\mathbf{\underline{p_i}}$ is defined as below:
\begin{itemize}
\item $\mathbf{a_i}$ corresponds to the coding coefficients $\alpha_j$, $j \in I_i$, where $I_i \subseteq V$ is the set of nodes adjacent to $v_i$ in $E_1$,
\item $\mathbf{h_{I_i}}$ corresponds to the hash $h(x_j)$, $v_j \in I_i$ where $h(\cdot)$ is a $\delta$-bit polynomial hash function,
\item $\mathbf{h_{x_i}}$ corresponds to the polynomial hash $h(x_i)$,
\item $\mathbf{x_i}$ is the $n$-bit representation of $x_i = \sum_{j \in I} \alpha_j x_j$.
\end{itemize}

The payload $\mathbf{x_i}$ is coded with a $(n, k_i)$-code $\mathcal{C}_i$ with minimum distance $d_i$. Code $\mathcal{C}_i$ is an error-correcting code of rate $R_i = \frac{k_i}{n} = 1-\frac{d_i}{n}$, and is tailored for the forward communication. For instance, $v_{1}$ uses code $\mathcal{C}_1$, chosen appropriately for the channel $(v_1, v_j)\in E_1$, to transmit the payload $\mathbf{x_1}$.

We assume that the payload $\mathbf{x_i}$ is $n$-bits, and the hash $h(\cdot)$ is $\delta$-bits. We assume that the hash function used, $h(\cdot)$, is known to all nodes, including the adversary. In addition, we assume that $\mathbf{a_i}$, $\mathbf{h_{I_i}}$ and $\mathbf{h_{x_i}}$ are part of the header information, and are sufficiently coded to allow the nodes to correctly receive them even under noisy channel conditions. Note that the hashes $\mathbf{h_{I_i}}$ and $\mathbf{h_{x_i}}$ are contained within one hop, and the overhead associated with the hashes is proportional to the in-degree of a node, and does not accumulate with the routing path length.

Assume that $v_i$ transmits $\mathbf{\underline{p_i}}=[\mathbf{a_i}, \mathbf{h_{I_i}}, \mathbf{h_{x_i}}, \mathbf{\hat{x}_i}]$, where $\mathbf{\hat{x}_i} = \mathbf{x_i} \oplus \mathbf{e}$,  $\mathbf{e} \in \{0,1\}^n$. If $v_i$ is misbehaving, then $\mathbf{e} \ne 0$. Our goal is to probabilistically detect when $\mathbf{e} \ne 0$. Note that even if $|\mathbf{e}|$ is small
(\ie small Hamming distance between $\mathbf{\hat{x}_i}$ and $\mathbf{x_i}$),
the algebraic interpretation of $\mathbf{\hat{x}_i}$ and $\mathbf{x_i}$ may differ significantly.

\subsection{Threat Model}\label{sec:threat}

We assume powerful adversaries, who can eavesdrop its neighbor's transmissions, has the power to inject or corrupt packets, and are computationally unbounded. However, the adversary does not know the specific realization of the random errors introduced by the channels. The adversaries' objective is to corrupt the information flow without being detected by other nodes. Thus, the adversary will find $\mathbf{\hat{x}_i}$ that will allow its misbehavior to be undetected, if there is any such $\mathbf{\hat{x}_i}$.

Our goal is to detect probabilistically a malicious behavior that is beyond the channel noise, represented by $BSC(p_{ik})$. Note that the algebraic watchdog does not completely eliminate errors introduced by the adversaries; its objective is to limit the errors introduced by the adversaries to be at most that of the channel. Channel errors (or those introduced by adversaries below the channel noise level) can be corrected using appropriate error correction schemes, which will be necessary even without Byzantine adversaries in the network.

The notion that adversarial errors should sometimes be treated as channel noise has been introduced previously in \cite{milcom}. Under heavy attack,
 attacks should be treated with special attention; while under light attack, the attacks can be treated as noise and corrected using error-correction schemes. The results in this paper partially reiterate this idea.
\begin{figure}[tbp]
\begin{center}
\includegraphics[width=0.28\textwidth]{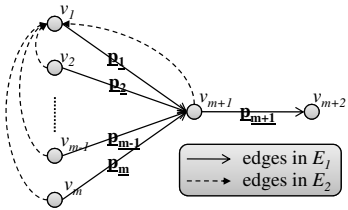}
\end{center}
\vspace*{-0.1cm}
\caption{A small neighborhood of a wireless network with $v_1$.}\label{fig:general}\vspace*{-0.5cm}
\end{figure}

\section{Algebraic Watchdog for Two-hop network}\label{sec:multisource}

Consider a small neighborhood of nodes in $G$ with nodes $v_1, v_2, ... v_m,$ $v_{m+1}$, $v_{m+2}$. Nodes $v_i$, $i\in [1,m]$, want to transmit $x_i$ to $v_{m+2}$ via $v_{m+1}$. Without complete information about all the messages, we cannot verify whether $v_{m+1}$ is misbehaving or not with certainty. However, there is a large overhead associated with acquiring complete information. Therefore, we take advantage of the wireless setting, in which nodes can overhear their neighbors' transmission (as shown in Figure \ref{fig:general}), to probabilistically detect malicious behavior. Each node checks whether its downstream neighbors are transmitting values that are consistent with the gathered information. If a node detects that its downstream neighbor is misbehaving, it can alert other nodes within the network.

The graphical model illustrates the inference process a node executes to check its next hop node. Without loss of generality, we consider the problem from $v_1$'s perspective. Denote $\tilde{x}_i$ to be the noisy message $v_1$ overhears from node $v_i$'s transmission, for any $i \in [2, m]$. Since the header is protected, $v_1$ correctly decodes $h(x_i)$ and $\alpha_i$ for all $i \in [1, m]$.

\subsection{Transition matrix}\label{sec:transitionprob}

We define a \emph{transition matrix} $T_i$ to be a $2^{n(1-H(\frac{d_i}{n}))+\delta}\times 2^{n(1-H(\frac{d_i}{n}))}$ matrix, where $H(\cdot)$ is the entropy function.
\begin{align*}
T_i(\tilde{x}_i, y) &=
\begin{cases}
\frac{p_i(\tilde{x}_i, y)}{\mathcal{N}}, &\text{if $h(y) = h(x_i)$}\\
0, &\text{otherwise}
\end{cases},\\
p_i(\tilde{x}_i, y) &= p_{i1}^{\Delta(\mathbf{\tilde{x}_i, y})}(1-p_{i1})^{n-\Delta(\mathbf{\tilde{x}_i, y})},\\
\mathcal{N} &= \sum_{\{y|h(y)=h(x_i)\}} p_i(\tilde{x}_i, y),
\end{align*}
where $\Delta(\mathbf{x}, \mathbf{y})$ gives the Hamming distance between codewords $\mathbf{x}$ and $\mathbf{y}$. In other words, $v_1$ computes $\tilde{X}_i = \{x|h(x) = h(x_i)\}$ to be the list of \emph{candidates} of $x_i$. For any overheard pair $\mathbf{[\tilde{x}_i, h(x_i)]}$, there are multiple candidates of $x_i$ (\ie $|\tilde{X}_i|$) although the probabilities associated with each inferred $x_i$ are different. This is because there are uncertainties associated with the wireless medium, represented by $BSC(p_{i1})$.

For each $x\in \tilde{X}_i$, $p_i(\tilde{x}_i, x)$ gives the probability of $x$ being the original codeword sent by node $v_i$ given that $v_1$ overheard $\tilde{x}_i$ under $BSC(p_{i1})$. Since we are only considering $x\in \tilde{X}_i$, we normalize the probabilities using $\mathcal{N}$ to get the \emph{transition probability} $T_i(\tilde{x}_i, x)$. Note $T_i(\tilde{x}_i, y)=0$ if $h(y) \ne h(x_i)$.

\begin{figure*}[tbp]
\begin{center}\hspace*{-.5cm}
\includegraphics[width=1.05\textwidth]{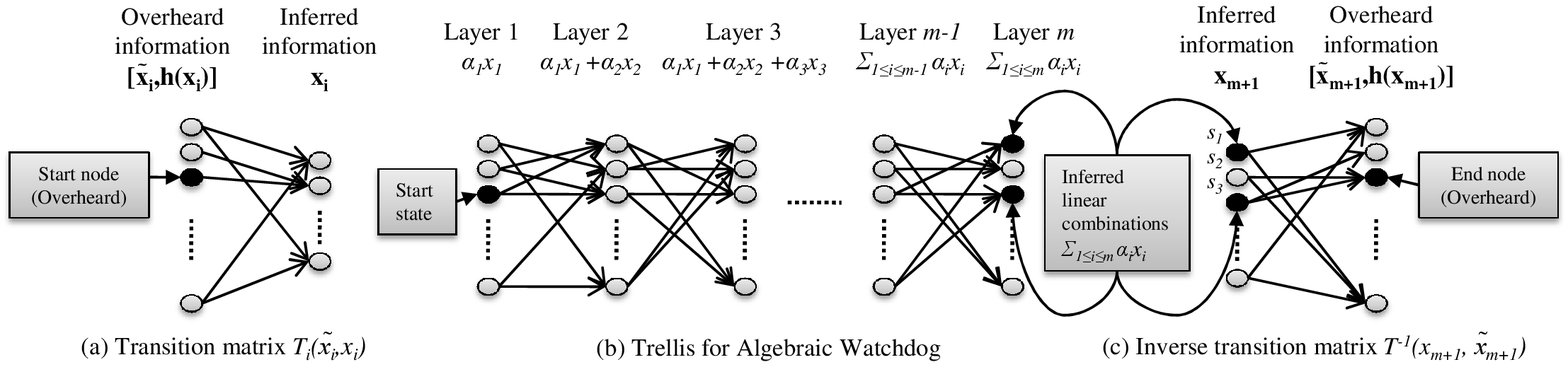}
\end{center}\vspace*{-.3cm}
\caption{Graphical representation of the inference process at node $v_1$. In the trellis, the transition probability from Layer $i-1$ to Layer $i$ is given by $T_i(\tilde{x}_i,x_i)$, which is shown in (a).}\vspace*{-.3cm}\label{fig:inference}\label{fig:trellis}\label{fig:transprob}\label{fig:invtrans}
\end{figure*}

The structure of $T_i$ heavily depends on the collisions of the hash function $h(\cdot)$ in use. Note that the structure of $T_i$ is independent of $i$, and therefore, a single transition matrix $T$ can be precomputed for all $i \in [1, m]$ given the hash function $h(\cdot)$. A graphical representation of $T$ is shown in Figure \ref{fig:transprob}a. For simplicity of notation, we represent $T$ as a matrix; however, the transition probabilities can be computed efficiently using hash collision lists as well.

\subsection{Watchdog trellis}\label{sec:trellis}

Node $v_1$ uses the information gathered to generate a trellis, which is used to infer the valid linear combination that $v_{m+1}$ should transmit to $v_{m+2}$. As shown in Figure \ref{fig:trellis}b, the trellis has $m$ layers: each layer may contain up to $2^n$ states, each representing the inferred linear combination so far. For example, Layer $i$ consist of all possible values of $\sum_{j=1}^i \alpha_j x_j$.

The matrices $T_i, i\in[2,m]$, defines the connectivity of the trellis.
 Let $s_1$ and $s_2$ be states in Layer $i-1$ and Layer $i$, respectively. Then, an edge $(s_1, s_2)$ exists if and only if
\[
\exists\ x \text{ such that } s_1 + \alpha_i x = s_2,\ T_i(\tilde{x}_i, x) \ne 0.
\]
We denote $w_e(\cdot, \cdot)$ to be the edge weight, where $w_e(s_1, s_2)=T_i(\tilde{x}_i, x)$ if edge $(s_1, s_2)$ exists, and zero otherwise.

\subsection{Viterbi-like algorithm}\label{sec:viterbi}

We denote $w(s,i)$ to be the weight of state $s$ in Layer $i$. Node $v_1$ selects a \emph{start state} in Layer 1 corresponding to $\alpha_1 x_1$, as shown in Figure \ref{fig:trellis}. The weight of Layer 1 state is $w(s, 1) = 1$ if $s = \alpha_1 x_1$, zero otherwise. For the subsequent layers, multiple paths can lead to a given state, and the algorithm keeps the aggregate probability of reaching that state. To be more precise, $w(s, i)$ is:
\[
w(s, i) = \sum_{\forall s' \in \text{Layer } i-1} w(s', i-1)\cdot w_e(s',s).
\]
By definition, $w(s,i)$ is equal to the total probability of $s = \sum_{j=1}^i \alpha_j x_j$ given the overheard information. Therefore, $w(s, m)$ gives the probability that $s$ is the valid linear combination that $v_{m+1}$ should transmit to $v_{m+2}$. It is important note that $w(s,m)$ is dependent on the channel statistics, as well as the overheard information. For some states $s$, $w(s,m) =0$, which indicates that state $s$ can not be a valid linear combination; only those states $s$ with $w(s,m)>0$ are the \emph{inferred candidate linear combinations}.

Note that the algorithm introduced above is a dynamic program, and is similar to the Viterbi algorithm. Therefore, tools developed for dynamic programming/Viterbi algorithm can be used to compute the probabilities efficiently.

\subsection{Decision making}\label{sec:decision}

Node $v_1$ computes the probability that the overheard $\tilde{x}_{m+1}$ and $h(x_{m+1})$ are consistent with the inferred $w(\cdot, m)$ to make a decision regarding $v_{m+1}$'s behavior. To do so, $v_1$ constructs an \emph{inverse transition matrix} $T^{-1}$, which is a $2^{n(1-\frac{d_{m+1}}{n})} \times 2^{n(1-\frac{d_{m+1}}{n})+\delta}$ matrix whose elements are defined as follows:
\begin{align*}
T^{-1}(y, \tilde{x}_{m+1}) &=
\begin{cases}
\frac{p_{m+1}(\tilde{x}_{m+1}, y)}{\mathcal{M}}, &\text{if $h(y) = h(x_{m+1})$}\\
0, &\text{otherwise}
\end{cases},\\
\mathcal{M} &= \sum_{\{y|h(y)=h(x_{m+1})\}} p_{m+1}(\tilde{x}_{m+1}, y).
\end{align*}
Unlike $T$ introduced in Section \ref{sec:transitionprob}, $T^{-1}(x, \tilde{x}_{m+1})$ gives the probability of overhearing $[\tilde{x}_{m+1}, h(x_{m+1})]$ given that $x \in \{y|h(y)=h(x_{m+1})\}$ is the original codeword sent by $v_{m+1}$ and the channel statistics. Note that $T^{-1}$ is identical to $T$ except for the normalizing factor $\mathcal{M}$. A graphical representation of $T^{-1}$ is shown in Figure \ref{fig:invtrans}c.

In Figure \ref{fig:invtrans}c, $s_1$ and $s_3$ are the inferred candidate linear combinations, \ie $w(s_1, m)\ne 0$ and $w(s_2,m)\ne 0$; the \emph{end node} indicates what node $v_1$ has overheard from $v_{m+1}$. Note that although $s_1$ is one of the inferred linear combinations, $s_1$ is not connected to the end node. This is because $h(s_1) \ne h(x_{m+1})$. On the other hand, $h(s_2) = h(x_{m+1})$; as a result, $s_2$ is connected to the end node although $w(s_2, m)= 0$. We define an inferred linear combination $s$ as \emph{matched} if $w(s,m)>0$ and $h(s) = h(x_{m+1})$.

Node $v_1$ uses $T^{-1}$ to compute the total probability $p^*$ of hearing $[\tilde{x}_{m+1}, h(x_{m+1})]$ given the inferred linear combinations by computing the following equation:
\[
p^* = \sum_{\forall s} w(s, m)\cdot T^{-1}(s, \tilde{x}_{m+1}).
\]
Probability $p^*$ is the probability of overhearing $\tilde{x}_{m+1}$ given the channel statistics; thus, measures the likelihood that $v_{m+1}$ is consistent with the information gathered by $v_1$. Node $v_1$ can use $p^*$ to make a decision on $v_{m+1}$'s behavior. For example, $v_1$ can use a threshold decision rule to decide whether $v_{m+1}$ is misbehaving or not: $v_1$ claims that $v_{m+1}$ is malicious if $p^* \leq t$ where $t$ is a threshold value determined by the given channel statistics; 
otherwise, $v_1$ claims $v_{m+1}$ is well-behaving.

Depending on the decision policy used, we can use the hypothesis testing framework to analyze the probability of false positive and false negative. Reference \cite{algebraicwatchdog} provides such analysis for the simple two-hop network. However, the purpose of this paper is not to propose a decision policy, but to propose a method in which we can compute $p^*$, which can be used to establish trust within a network. We note that it would be worthwhile to look into specific decision policies and their performance (\ie false positive/negative probabilities) as in \cite{algebraicwatchdog}.

\section{Analysis for Two-hop Network}\label{sec:analysis}

In this section, we provide an analysis for the performance of Algebraic Watchdog for two-hop network.

\begin{theorem}\label{thm:number}
Consider a two-hop network as shown in Figure \ref{fig:general}. Consider node $v_j$, $j \in [1, m]$. Then, the number of \emph{matched} codewords is:
\[
2^{n\left[\sum_{i \ne j, i\in[1, m+1]} \left(H(p_{ij})-H(\frac{d_i}{n})\right) -1 \right] - m\delta}.
\]
\end{theorem}
\begin{proof}
Without loss of generality, we consider node $v_1$. The proof uses on concepts and techniques developed for list-decoding \cite{listdecoding}. We first consider the overhearing of $v_k$'s transmission, $k \in [2, m]$. Node $v_1$ overhears $\tilde{x}_k$ from $v_k$. The noise introduced by the overhearing channel is characterized by $BSC(p_{k1})$; thus, $E[\Delta(\mathbf{x_k, \tilde{x}_k})] = np_{k1}$. Now, we consider the number of codewords that are within $B(\tilde{x}_k, np_{k1})$, the Hamming ball of radius $np_{k1}$ centered at $\tilde{x}_k$ is
$|B(\tilde{x}_k, np_{k1})| = 2^{n(H(p_{k1})-H(\frac{d_k}{n}))}$.
Note that $v_1$ overhears the hash $h(x_k)$; thus, the number of codewords that $v_1$ considers is reduced to $2^{n(H(p_{k1})-H(\frac{d_k}{n}))-\delta}$.
Using this information, $v_1$ computes the set of inferred linear combinations, \ie $s$ where $w(s, m) >0$. Note that $v_1$ knows precisely the values of $x_1$. Therefore, the number of inferred linear combinations is upper bounded by:
\begin{align}
\prod_{k\in [2, m]}&\left(2^{n\left(H(p_{k1})-H(\frac{d_k}{n})\right)-\delta}\right)
\\ &= 2^{n\left[\sum_{k\in [2,m]}\left( H(p_{k1}) - H(\frac{d_k}{n})\right)\right] - (m-1)\delta} \label{eq:num_inferred}
\end{align}
Note that due to the finite field operations, these inferred linear combinations are randomly distributed over the space $\{0,1\}^n$.

Now, we consider the overheard information, $\tilde{x}_{m+1}$ from the downstream node $v_{m+1}$. By similar analysis as above, we can derive that there are $2^{n(H(p_{m+1, 1})-H(\frac{d_{m+1}}{n}))-\delta}$
codewords in the hamming ball $B(\tilde{x}_{m+1}, np_{m+1,1})$ with hash value $h(x_{m+1})$. Thus, the probability that a randomly chosen codeword in the space of $\{0,1\}^n$ is in $B(\tilde{x}_{m+1}, np_{m+1,1}) \cap \{x| h(x) = h(x_{m+1})\}$ is:
\begin{equation}\label{eq:num_relay}
\frac{2^{n(H(p_{m+1, 1})-H(\frac{d_{m+1}}{n}))-\delta}}{2^n}.
\end{equation}

Then, the expected number of \emph{matched} codewords is the product of Equations (\ref{eq:num_inferred}) and (\ref{eq:num_relay}).
\end{proof}

Note that if we assume that the hash is of length $\delta= \varepsilon n$, then the statement in Theorem \ref{thm:number} is equal to:
\begin{equation}
2^{n\left[\sum_{i \ne j, i\in[1, m+1]} H(p_{ij})- \left( \sum_{i \ne j, i\in[1, m+1]} H(\frac{d_i}{n}) +1  + m\varepsilon \right)\right]}.
\end{equation}
This highlights the tradeoff between the quality of overhearing channel and the redundancy (introduced by $\mathcal{C}_i$'s and the hash $h$). If enough redundancy is introduced, then $\mathcal{C}_i$ and $h$ together form an error-correcting code for the overhearing channels; thus, allows exact decoding to a single matched codeword.

The analysis also shows how adversarial errors can be interpreted. Assume that $v_{m+1}$ wants to inject errors at rate $p_{adv}$. Then, node $v_1$, although has an overhearing $BSC(p_{m+1,1})$, effectively experiences an error rate of $p_{adv} + p_{m+1,1} - p_{adv} \cdot p_{m+1, 1}$. Note that this does not change the set of the inferred linear combinations; but it affects $\tilde{x}_{m+1}$. Thus, overall, adversarial errors affect the set of matched codewords and the distribution of $p^*$. As we shall see in Section \ref{sec:simulation}, the difference in distribution of $p^*$ between a well-behaving relay and adversarial relay can be used to detect malicious behavior.

\section{Protocol for Algebraic Watchdog}\label{sec:multihop}

In this section, we use the two-hop algebraic watchdog from Section \ref{sec:multisource} in a hop-by-hop manner to ensure a globally secure network. In Algorithm \ref{alg:protocol}, we present a distributed algorithm for nodes to secure the their local neighborhood. Each node $v$ transmits/receives data as scheduled; however, node $v$ randomly chooses to check its neighborhood, at which point node $v$ listens to neighbors transmissions to perform the two-hop algebraic watchdog from Section \ref{sec:multisource}.

\begin{algorithm}[b]
\vspace*{-.5cm}
\ForEach{node $v$}{
According to the schedule, transmit and receive data;
\If{$v$ decides to check its neighborhood}
{
Listen to neighbors' transmissions;\\
\ForEach{downstream neighbor $v'$}
{
Perform Two-hop Algebraic Watchdog on $v'$;
}}}
\caption{Distributed algebraic watchdog at $v$.}\label{alg:protocol}
\end{algorithm}

\begin{corollary}\label{thm:v}
Consider $v_{m+1}$ as shown in Figure \ref{fig:general}. Assume that the downstream node $v_{m+2}$ is well-behaving, and thus, forces $\mathbf{h_{x_{m+1}}} = h(x_{m+1})$. Let $\mathbf{\underline{p_i}}$ be the packet received by $v_{m+1}$ from parent node $v_i \in P(v)$. Then, if there exists at least one well-behaving parent $v_j \in P(v)$, $v_{m+1}$ cannot inject errors beyond the overhearing channel noise ($p_{m+1,j}$) without being detected.
\end{corollary}

Section \ref{sec:analysis} shows that presence of adversarial error (at a rate above the channel noise) can be detected by a change in distribution of $p^*$. Note that this Corollary \ref{thm:v} does not make any assumptions on whether packets $\mathbf{\underline{p_i}}$'s are valid or not. Instead, the claim states that $v_{m+1}$ transmits a valid packet \emph{given} the packets $\mathbf{\underline{p_i}}$ it has received.

\begin{corollary}\label{thm:v2} Node $v$ can inject errors beyond the channel noise only if either of the two conditions are satisfied:
\begin{enumerate}
\item All its parent nodes $P(v) = \{u|(u,v)\in E_1\}$ are colluding Byzantine nodes;
\item All its downstream nodes, \ie receivers of the transmission $\mathbf{\underline{p_i}}$, are colluding Byzantine nodes.
\end{enumerate}
\end{corollary}

{\it Remark:} Note that, in Case 1) of Corollary \ref{thm:v2}, $v$ is not responsible to any well-behaving nodes. Node $v$ can transmit any packet without the risk of being detected by any well-behaving parent node. However, the min-cut to $v$ is dominated by adversaries, and the information flow through $v$ is completely compromised -- regardless of whether $v$ is malicious or not. In Case 2) of the Corollary \ref{thm:v2}, $v$ can generate any hash value since its downstream nodes are colluding adversaries. Thus, it is not liable to transmit a consistent hash, which is necessary for $v$'s parent nodes to monitor $v$'s behavior. However, note that $v$ is not responsible in delivering any data to a well-behaving node. Even if $v$ were well-behaving, it cannot reach any well-behaving node without going through a malicious node in the next hop. Thus, the information flow through $v$ is again completely compromised.

Therefore, Corollary \ref{thm:v2} shows that algebraic watchdog can aid in ensuring correct delivery of data when the following assumption holds: for every intermediate node $v$ in the path between source to destination, $v$ has at least one well-behaving parent and at least one well-behaving child -- \ie there exists at least a path of well-behaving nodes. This is not a trivial result as we are not only considering a single-path network, but also multi-hop, multi-path network.

\section{Simulations}\label{sec:simulation}

We present preliminary MATLAB simulation results that show the difference in distribution of $p^*$ between the well-behaving and adversarial relay. We consider a setup in Figure \ref{fig:general}. We set all $p_{i1}$, $i \in [2, m]$ to be equal, and we denote this probability as $p_s = p_{i1}$ for all $i$. We denote $p_{adv}$ to be the probability at which the adversary injects error; thus, the effective error that $v_1$ observes from an adversarial relay is combined effect of $p_{m+1, 1}$ and $p_{adv}$. The hash function $h(x) = ax + b \mod 2^\delta$ is randomly chosen over $a, b \in \mathbf{F}_{2}^\delta$. We set $n = 10$; thus, the coding field size is $2^{10}$. For each data point, we run the algebraic watchdog 200 times.

For simplicity, we assume that the nodes do not use an error-correcting code $\mathcal{C}_i$; thus, $d_i = 0$ for all $i$. Note that this limits the power of the algebraic watchdog; thus, the results shown can be further improved by using error correcting codes $\mathcal{C}_i$.

We denote $p^*_{adv}$ and $p^*_{relay}$ as the value of $p^*$ when the relay is adversarial and is well-behaving, respectively. We denote $var_{adv}$ and $var_{relay}$ to be the variance of $p^*_{adv}$ and $p^*_{relay}$.

\begin{table}[tbp]
\caption{The average and variance of $p^*$ with varying $p_{adv}$. We set $m = 3$, $n = 10$, $\delta = 2$, and $p_s = p_{m+1, 1} = 10\%$.}\label{tb:padv}
\centering
\begin{tabular}{|c|c|c|c|c|}
\hline
$p_{adv}$ & $p^*_{adv}$ & $var_{adv}$ & $p^*_{relay}$ & $var_{relay}$\\
\hline
0\% & 0.0262 & 0.0019 & 0.0262 & 0.0019 \\
5\% & 0.0205 & 0.0019 & 0.0259 &  0.0022\\
10\% & 0.0096 & $2.8933\times 10^{-4}$ & 0.0220 &  0.0012\\
15\% & 0.0129 & $7.7838 \times 10^{-4}$ & 0.0302 & 0.0023\\
20\% & 0.0139 & $7.7535 \times 10^{-4}$ & 0.0287 & 0.0018\\
30\% & 0.0093 & $5.6885 \times 10^{-4}$ & 0.0243 & 0.0012\\
\hline
\end{tabular}
\end{table}
\begin{table}[tbp]
\caption{The average and variance of $p^*$ with varying $\delta$. We set $m = 3$, $n = 10$, $p_s = p_{m+1, 1} = 10\%$, and $p_{adv} = 10\%$.}\label{tb:delta}
\centering
\begin{tabular}{|c|c|c|c|c|}
\hline
$\delta$ & $p^*_{adv}$ & $var_{adv}$ & $p^*_{relay}$ & $var_{relay}$\\
\hline
0& 0.0046 & $3.8552\times 10^{-4}$ & 0.0071 & $4.4103\times 10^{-4} $\\
1 & 0.0083 & $3.1523\times 10^{-4}$ & 0.0120 & $4.6351 \times 10^{-4}$\\
2 & 0.0096 & $2.8933\times 10^{-4}$ & 0.0220 &  0.0012\\
3 & 0.0067 & $2.3491\times 10^{-4}$ & 0.0240 & 0.0015\\
\hline
\end{tabular}\vspace*{-.4cm}
\end{table}

\begin{table}[t]
\caption{The average and variance of $p^*$ with varying $p_{s}$. We set $m = 3$, $n = 10$, $\delta = 2$, and $p_{adv} = 10\%$.}\label{tb:ps}
\centering
\begin{tabular}{|c|c|c|c|c|}
\hline
$p_{s}$ & $p^*_{adv}$ & $var_{adv}$ & $p^*_{relay}$ & $var_{relay}$\\
\hline
5\% & 0.0067 & $6.3986\times 10^{-4}$ & 0.0455 &  0.0051\\
10\% & 0.0096 & $2.8933\times 10^{-4}$ & 0.0220 &  0.0012\\
20\% & 0.0033 & $4.9045 \times 10^{-5}$ & 0.0055 & $7.3819 \times 10^{-5}$\\
30\% & 0.0069 & $1.6414 \times 10^{-4}$ & 0.0100 & $8.5259 \times 10^{-4}$\\
\hline
\end{tabular}
\end{table}
\begin{table}[tbp]
\caption{The average and variance of $p^*$ with varying $m$. We set $n = 10$, $\delta = 2$, $p_s = p_{m+1, 1} = 10\%$, and $p_{adv} = 10\%$.}\label{tb:m}
\centering
\begin{tabular}{|c|c|c|c|c|}
\hline
$m$ & $p^*_{adv}$ & $var_{adv}$ & $p^*_{relay}$ & $var_{relay}$\\
\hline
1& 0.0155 & 0.0025 & 0.1386 & 0.0238 \\
2 & 0.0087 & $1.2932\times 10^{-4}$ & 0.0225 &  0.0011\\
3 & 0.0096 & $2.8933\times 10^{-4}$ & 0.0220 &  0.0012\\
4 & 0.0082 & $1.1902\times 10^{-4}$ & 0.0136 & $3.4301\times 10^{-4}$\\
5 & 0.0063 & $4.1841 \times 10^{-5}$ & 0.0079 & $5.2244 \times 10^{-5}$\\
\hline
\end{tabular}\vspace*{-.3cm}
\end{table}

Our simulation results coincide with our analysis and intuition. Section \ref{sec:analysis} noted that $v_1$ can detect adversarial errors $p_{adv} \geq p_{m+1, 1}$. As shown in Table \ref{tb:padv}, node $v_1$ is able to monitor its downstream node $v_{m+1}$ when $p_{adv} \geq p_{m+1, 1}$. There is a significant change in the distribution of $p^*$ once $p_{adv} \geq p_{m+1, 1} = 10\%$. Table \ref{tb:delta} shows that increase in redundancy (by using hash functions of length $\delta$) helps $v_1$ detect malicious behaviors. Note that for this simulation, $\delta > 1$ gives enough redundancy for $v_1$ monitor $v_{m+1}$.

Table \ref{tb:ps} shows that overhearing channel between $v_1$ and $v_i$, $i \in [2, m]$ is important in detection. This agrees with our intuition -- if node $v_1$ is able to infer better the messages $x_{i}$, the better its detection abilities. Thus, as the overhearing channel progressively worsens ($p_s$ increase), $v_1$'s ability to detect malicious behavior deteriorates, as shown in Table \ref{tb:ps}.

In Table \ref{tb:m}, we note the effect of $m$. Node $v_1$'s ability to check $v_{m+1}$ is reduced with $m$. When $m$ increases, the number of messages to infer increases, which increases the uncertainty in the system. However, Table \ref{tb:m} does not take into account that with increase in $m$, there are more $v_i$'s, $i \in [1, m]$ that perform checks on $v_{m+1}$ independently.

\section{Conclusions}\label{sec:conclusions}

In this paper, we have proposed a multi-hop, multi-source algebraic watchdog, which allows network coded wireless systems to validate its information flows. A node monitors its downstream node by overhearing transmissions of its neighboring nodes; and uses the overheard information to infer what the behavior of its downstream node should be. Using the algebraic watchdog scheme, nodes can compute a probability of misbehavior, which can be used to detect malicious behavior. Once a node has been identified as malicious, these nodes can either be punished/eliminated or excluded from the network by using reputation based schemes \cite{marti}\cite{reputation}.

We have provided a trellis-like graphical model for the detection inference process, and provided an algorithm that may be used to compute the probability that a downstream node is consistent with the overheard information. We have analytically shown how the size of hash function, minimum distance of the code used, as well as the overhearing channel quality can affect the probability of detection. Finally, we have presented preliminary simulation results that coincide with our analysis and intuition.

\bibliography{References}
\bibliographystyle{IEEEtran}

\end{document}